\documentclass[11pt]{article}
\usepackage{graphicx}
\usepackage{hyperref}
\usepackage{verbatim}
\usepackage{amsmath, amsthm, amssymb, multicol, color}
\usepackage{float}
\usepackage{cite}
\usepackage{lscape}
\usepackage{listing}
\usepackage{subcaption}
\usepackage{mwe}
\usepackage{tikz}
\usetikzlibrary{matrix}
\usepackage{titlesec}
\usepackage[hmargin=1in,vmargin=1in]{geometry}
\usepackage{setspace}
\title{A General Approach to Coding in Early Olfactory and Visual Neural Populations} %or Early Visual and Olfactory Population Coding 
\begin{document}

\section*{\centering A General Approach to Coding in Early Olfactory and Visual Neural Populations}

\centering
William T. Redman

\centering
University of California, Santa Barbara

\centering
wredman@ucsb.edu

\abstract{Recent experimental and theoretical work on neural populations belonging to two separate early sensory systems, olfaction and vision, has challenged the notion that the two operate under different computational paradigms by providing evidence for the respective neural population codes having three central, common features: they are highly redundant; they are organized such that information is carried in the identity, and not the relative timing, of the active neurons; they are capable of error correction. We present the first model that captures these three properties in a general manner, making it possible to investigate whether similar structure is present in other population codes. Our model also makes specific predictions about additional, as yet unseen, structure in such codes. If these predictions are found in real data, this would provide new evidence that such population codes are operating under more general computational principles. }

\section*{Introduction}

%The coding properties of early sensory systems have been the focus of a large body of experimental and theoretical literature, because of their relative ease to record from and their importance in being the input to higher level brain areas. Examples of topics addressed are representation of sequences in the retinal population code and predictive coding ([xx]), pairwise correlations in the retinal code ([xx]),  (NEED OLFACTORY LITERATURE)

%How early sensory systems encode information about the world at the neural level has a long history in the literature because of the implications of understanding this coding for medical applications, as well as the hope that understanding how the neural code is organized at the ``lower'' levels will inform how the neural code is organized at ``higher'' levels.  

%Biological and neural systems are notoriously noisy and yet, in many cases, highly robust. The difficulty in reconciling these two aspects of biological and neural systems has led to a whole body of work (for just a few of many examples see \cite{Wolpert} - \cite{Desplan}), which has resulted in insights as to how networks with interacting components can have robust output in the presence of noise. 

Because of their relative ease to record from and their importance in being the input to upstream brain areas, the coding properties of neural populations belonging to early sensory systems have been the focus of a large body of experimental and theoretical literature. This work has explored the precision of retinal spike trains \cite{Berry}, the correlation structure of individual retinal ganglion cells and their population wide collective states \cite{Schneidman}, \cite{Tkacik2}, the possibility of criticality in retinal populations \cite{Tkacik}, \cite{Mora}, the preservation of odor identity representation by glomeruli (structures in the olfactory bulb that receive projections from olfactory sensory neurons - OSNs) across varying odor concentrations \cite{Gross-Isseroff}, \cite{Bhagavan}, \cite{Uchida}, \cite{Cleland} and the combinatorial nature of the glomeruli code \cite{Malnic}, \cite{Saito}. While the respective work has followed a similar mission (to understand the nature of the population codes), the results have caused a divergence in the belief of the coding principles at use. Yet, in both systems, the exact structure of the exact nature of the population code has been unclear. 
%Despite this breadth and depth of work, a number of questions remain, including the exact structure of the population codes and how they are processed by higher brain regions. 

Recent work has taken advantage of advances in recording and manipulation technology, as well as statistical methods, to probe more intricately at this question of the exact structure of the retinal ganglion and the glomeruli population codes \cite{Prentice}, \cite{Loback}, \cite{Wilson}, \cite{Giaffar}. Despite the differences in system and approach, the two have converged on three similar principles for the respective codes. First, the codes are highly redundant, as only a small subset of the neurons are responsible for carrying the information about the identity of any given stimulus (whether it be some feature of the visual field or an odor). Second, it is the identity of these neurons in the specific subset (and not their relative timing) that carries the relevant information. Finally, a neural population code with those two features is endowed with the capability to be robust to noise, in the sense that the code is capable of error correction. These core similarities clearly challenge the assumption that the two systems are operating under different coding paradigms.

In the retinal ganglion population code, these features come about by the fact that the probability space of population responses is populated by geometric objects identified as ridges. These ridges correspond to unique ``codewords'' that a downstream system maps all responses in the ridge to \cite{Loback}. The identity of these ridges is determined by an active set of neurons (those neurons that were active in the states that make up a given ridge) and a silent set (those neurons that weren't active). The mapping from a given neural response to the appropriate, ridge-specific codeword was hypothesized to be achieved by an additional layer of neurons, each one firing if a certain fraction of neurons in a given active set fire and none of neurons in the corresponding silent set fire (see \cite{Loback} Fig. 13). Under this simple model, it's easy to see that there could be a number of population response states that get mapped to the same codeword. 

In the glomeruli population code, these features arise from the fact that the code has been found operate under the ``Primacy Hypothesis'', namely that it is the first $n$ active glomeruli that are responsible for encoding odor identity \cite{Wilson}. Therefore, all population responses that have the same first $n$ active glomeruli (where the order of the first $n$ active glomeruli doesn't matter \cite{Wilson}), are recognized as being the same odor. Any of those given states can therefore be seen as a codeword. The relevant time scale (which determines the relevant $n$) was found to be $< 100$ ms \cite{Wilson}. Whether $n$ is fixed, or whether it can vary for various odors and odor mixtures, has yet to be determined. 

Recent theoretical work has suggested that the aforementioned three properties might be a more universal feature of neural population codes \cite{Ioffe}. In particular, brain regions such as MT and V1, which have similar firing rates and pairwise correlations to the retinal ganglion cells, were hypothesized to have these same properties in their neural codes. A more general framework in which to talk about neural population codes with these three properties is therefore desirable, especially if it can make predictions to specific structures that these codes should have. 

The rest of this paper will focus on such a framework. In the \textbf{Method} section, we outline our model for mapping arbitrary neural population responses to codewords. Our model utilizes the universal property of free groups (UPFG). We then provide an example of our model by mapping a specific set of neural responses to a given set codewords. We compare our methods mapping with the mapping generated by using Generalized Minimum Distance (GMD) decoding \cite{Forney}, a well known method in coding theory. This example is given to provide a more intuitive feel for our model (especially for those unfamiliar with group theory) and illustrate how it compares to other known (but not directly implicated in neural population coding) error correction algorithms. Finally, in the \textbf{Discussion} section, we discuss the implications of our model with regards to real neural population data and possible future directions for this work.

\section*{Why use the UPFG?}

There are two reasons why we turn to using such a formal, and mathematically abstract, language as group theory for this problem. First, for those with a background in group theory, we believe that our model is fairly intuitive. Second, and more pertinently, theoretical neuroscience has seen great advances when \textit{appropriate} theories/descriptions from formal physics and mathematics have been applied to neural problems. For instance, by  borrowing ideas and analysis techniques from statistical physics, attractor neural networks (ANNs) were able to be developed and thoroughly explored  \cite{Hopfield}, \cite{Amit}. In this vein, we feel that when considering redundant neural codes that ``collapse'' different population response states onto the same output (or recognized) state, the theory associated with homomorphisms (and group theory), is an appropriate language to use. For those unfamiliar with group theory, see some basic discussion in the \textbf{Material and Methods} section for a brief discussion of the simple group theoretic definitions and notions used in our model.

\section*{Model}

We start by defining the \textit{set of generators}, $G$, as $G = S + B$, where $S = \{1,2,...,n\}$, $B$ is a subset of the power set of $S$ (i.e. $B \subset \mathcal{P}(S)$), and $+$ is the set concatenation operator. We refer to $B$ as the \textit{basis set}. The relevance of $B$ will be discussed below. 

A simple example of this is, for $n = 5$, $G = \{1,2,3,4,5, 12, 145\}$, where the last two elements are the elements of the basis set (here $12$ stands for $\{1, 2\}$). 

We define the \textit{group of codewords}, $(C, *_C)$, as $C = \{c_1, c_2, ... ,c_m\}$ and $*_C$ is some operation on the elements of $C$ that meets the standard group criteria \cite{Dummit}. 

Finally, the \textit{restricted free group of G}, $\tilde{F}(G)$, is defined to be the abeleanized group of all elements in the free group of $G$, $F(G)$, that are made up of, at most, each element of $G$ once. For instance, while $1 2 1 \in F(G)$, $1 2 1  \notin \tilde{F}(G)$ because it has $1$ twice.

Using the universal property of free groups \cite{Dummit}, we have the following diagram, 

\begin{equation}
\begin{tikzpicture}
  \matrix (m) [matrix of math nodes,row sep=3em,column sep=4em,minimum width=2em]
  {
  G & F(G) \\
   & C\\};
   \path[-stealth]
    (m-1-1) edge node [below] {$i$} (m-1-2)
    (m-1-1) edge node [left] {$g$} (m-2-2)
    (m-1-2) edge node [right] {$\varphi$} (m-2-2);
  
\end{tikzpicture}  
\end{equation}
where $i$ is the inclusion map (i.e. $i(x) = x$ for all $x \in G$), $g$ is a group function determining $C$ from $G$, and $\varphi$ is a unique homomorphism (given a specific $g$) from $F(G)$ to $C$. The UPFG tells us that we can relate $g$ to $\varphi$ by 
\begin{equation}
\varphi(A) = g(a_1) *_C ... *_C g(a_n)
\end{equation}
where $A \in F(G)$ and $A = a_1 ... a_n$, such that $a_i \in G$ for all $1 \leq i \leq n$. 

A final point on our model must be made. We define, for all $s \in S$,
\begin{equation}
g(s) = \text{id}_C
\end{equation}
where id$_C$ is the identity element of the group of code words (i.e. for all $a \in C$, $a \hspace{1 mm} *_C$ id$_C = a$).

With this definition, it is more clear what the role of the basis set, $B$, is (as it is unnecessary for ``building" $F(G)$); it is the generator of $(C,*_C)$ under $g$. 

We now consider an element (or ``word''), $\omega$, of $\tilde{F}(G)$ (we consider $\tilde{F}(G)$ because of it represents all possible neural responses that were considered in the neural data from \cite{Prentice} - \cite{Giaffar}, but the same holds true, with small modifications, for any element in $F(G)$). We want to find the codeword (i.e. the element in $(C, *_C)$ that corresponds to $\omega$. In particular, for any given $\omega$, there exists a subset of $B$, $\{b_1,...,b_k\}$, and a subset of $S$, $\{s_1,..,s_m\}$ ($s_j \in S$), s.t. 

\begin{equation}
\omega = b_1 ... b_k s_1 ... s_m
\end{equation}
(the exact ordering of the $b_i$'s and $s_j$'s doesn't matter because $\tilde{F}(G)$ is abelian). Of course, this decomposition of $\omega$ is by no means unique. We therefore define the decomposition of $\omega$ as the pair $(\{ b_i \}, \{s_j\})$, s.t. the number of elements of $\{ b_i \}$ is the (if possible) non-zero minimum of all the possible decompositions of $\omega$. From the neural perspective, this is equivalent to demanding that every word is represented as simply as possible by the activity states that make up the basis set. 

With this, we can now look at the mapping of $\omega$ to its relevant codeword, which is given by $\varphi(\omega)$ (since $\varphi: \tilde{F}(G) \rightarrow C$). 

\begin{equation}
\varphi(\omega) = \varphi(b_1 ... b_k s_1 ... s_l) 
\end{equation}
$$
= g(b_1) *_C ... *_C g(b_k) *_C g(s_1) *_C ... *_C g(s_l) 
$$
$$
= g(b_1) *_C ... *_C g(b_k)
$$
By the defined decomposition of $\omega$, this decoding is unique. 

Note therefore that if two elements, $\omega_1$ and $\omega_2$, in $\tilde{F}(G)$ have the same basis elements, $\{b_i\}$, in their decomposition, then the decoding of the two elements is equivalent 
\begin{equation}
\varphi(\omega_1) = g(b_1) *_C ... *_C g(b_k) = \varphi(\omega_2)
\end{equation}

\subsection*{Example}
To illustrate our model, we map an example neural response space onto to example codewords. We also compare this mapping to an existing error correction method, Generalized Minimal Distance (GMD) decoding \cite{Forney}, as a way to show the possible strengths of our method in reference to existing methods. 

We will consider $G= \{1,2,3,4, 1 2, 2 4\}$, $C = \big(\{ 0, 1 2, 2 4, 1 4\}, +\big)$. For simplicity, we will convert each element of $G$ and $C$ into a binary string. This corresponds to $G = \{1000, 0100, 0010, 0001, 1100, 0101\}$ and $C = \big(\{0000, 1100, 0101, 1001\}, +$ $\text{mod(2)}\big)$, if we take each number, $1,..,4$, to be a position in a four bit string that has a value of $1$. 

The result of applying our mapping method and applying GMD decoding (where the decoding is determined by the codeword that has the minimal Hamming distance from the word we are trying to decode) is given in Table 1. Note that we are assuming that every word is equally likely to be received. From this, we see that, first and foremost, the number of three way ties (as denoted by $?$) is significantly less using the UPFG decoding as opposed to GMD decoding (one vs. eight). Additionally, in every determined decoding, the two methods agree.

\begin{table}[H]
\centering
\caption{Comparison of our model (UPFG mapping) and GMD decoding on an example response space}
\begin{tabular}{c|c|c}
Word & UPFG decoding & GMD decoding \\
\hline
1111 & ? & ?\\
1110 & 1100 & 1100\\
1101 & 1100 & ?\\
1011 & 1001 & 1001\\
0111 & 0101 & 0101\\
1100 & 1100 & 1100\\
1001 & 1001 & 1001\\
0011 & 0000 & ?\\
1010 & 0000 & ?\\
0101 & 0101 & 0101\\
0110 & 0000 & ?\\
1000 & 0000 & ?\\
0100 & 0000 & ?\\
0010 & 0000 & 0000\\
0001 & 0000 & ?\\
0000 & 0000 & 0000
\end{tabular}
\caption{}
\end{table}

\section*{Discussion}

Recent work investigating the structure of the retinal ganglion and glomeruli population codes, \cite{Prentice} - \cite{Giaffar}, has challenged a number of widely held theoretical biophysical and neural beliefs with their three central (and convergent) conclusions: the respective neural population codes are redundant, in the sense that multiple neural response states are interpreted as encoding the same information; it is the identity of the relevant subset of neurons that fire, and not the relative timing of their firing, that encodes the identity of the stimulus; such codes with the previous two properties are endowed with the capability of error correction. 

The fact that these neural population codes were found to be redundant contradicts the belief that neural codes should be very efficient. Such conclusions will force theorists to reconsider what, if anything, neural population codes are optimized for. Additionally, the finding that the first two properties allow for error correction marks a transition from the focus of biophysical and neural theory on how systems can output robustly despite noise (for a few of many examples, see \cite{Wolpert}, \cite{Gregor}, \cite{Elowitz}, \cite{Losick}) to the idea that the output of neural systems can be noisy, but it is the subsequent mapping (or ``interpretation'') of that output that is robust to noise.

Finally, that these conclusions are reached by very different experimental and theoretical methods in the retinal ganglion and glomeruli populations suggest that the two sensory systems (vision and olfaction), despite being previously believed to be operating under different coding principles, are, in fact, using the same principles. This surprising (and powerful) fact, coupled with theoretical work \cite{Ioffe} arguing that these coding principles might be used in more than just early sensory systems, and in higher brain areas like MT and V1, highlights the need for a general framework in which to discuss codes with such three properties. 

%Therefore, because our model has correspondences with each of the three aspects of the recent work, it's natural to look at whether there's a correspondence between some feature of the neural population codes and the basis set in our model. In the case of olfaction, there is a very obvious mapping between the basis set and the ``primacy'' set in (\cite{Primacy code}). Indeed, our model is more or less identical to the primacy coding model presented by Rinberg's group, the difference being that our is more general (and not specific to olfaction). 

To see that our model indeed is a general framework that meets all three of these facets of neural population codes, note first that it is only the basis elements that determine the decoding of any word. Therefore, words that vary by elements that don't affect the basis elements are seen as equivalent (i.e. eq. 6). This can also be clearly seen in Table 1 (e.g. 1100 and 1101 have the same mapping). Second, the fact that we restrict ourselves to looking specifically at the restricted free group of $G$, $\tilde{F}(G)$, we are not only restricting ourself to the more reasonable neural case where neurons are consider to have fired at most once in a time bin, we are also restricting ourselves to only the identity of the basis elements mattering in decoding and not the ordering, as $\tilde{F}(G)$ is abelian, and hence, the words $\omega_1 \omega_2$ and $\omega_2 \omega_1$ are equivalent. Finally, as noted before, because only the basis elements in the decomposition of a given word determine the decoding, our method is capable of error correction.  

Our model, while capturing the three main principles of the aforementioned work, also makes predictions about aspects of the early sensory systems' coding. In particular it predicts the existence of a basis set. In the context of the glomeruli population code, there exist two possibilities. Either one, $n$ (the number of relevant active glomeruli for odor classification) is fixed and the basis set is the set of codewords, which gives no new insight into the code. If $n$ is not fixed, then there exists the possibility of all codewords being able to be built from some smaller set of response states that are themselves codewords. For example, if $n = 2, 3,$ and $4$ are all allowed (e.g. $n$ is a function of odor complexity), then any $n = 4$ codeword could be built from two codewords that are $n = 2$. This greatly reduces the amount of elements needed for describing the codewords, and sheds light onto possible downstream decoding mechanisms. 

Similarly, for the retinal population code, the existence of a non-trivial basis set would shed light on downstream decoding schemes. In particular, it would provide possible adjustments to the model hypothesized in \cite{Loback} (i.e. Fig. 13). It is important to note that looking for the basis set in the retinal population code may require a switch in perspective, where it is the neurons in a given codewords silent set that might be the relevant feature. Future work will focus on methods for searching for basis sets, as well as specific implications of the existence of basis sets for downstream decoding schemes and how such schemes might develop in a natural way. Finding such basis sets in both the glomeruli and retinal population codes would extend even further the growing understanding of how similar the two codes are, and provide more evidence for the two operating under more universal computing principles. 

One clear failure of our model is that is relies on group theoretic notions that most in the neuroscience community aren't familiar with. While we believe that continuing to think in terms of this language will be useful (and indeed, we hope that our model convinces others in the mathematical community to consider the possible role free groups, and the UPFG, might play in error correction and neural coding), we hope to translate this model into a more clear, and less mathematically technical, model that still captures the main principles, but can more easily be communicated to others.

We hope that we have made clear the similarities of the two early sensory systems' population codes, the need for a general framework in which to explore arbitrary codes that share the properties exhibited by those two codes, and the possible utility of using group theory as a language to talk about such codes. 

\subsection*{Materials and Methods}

We provide here the basic group theory definitions and notions that are required for understanding our model. We have tried to make this as easy to understand as possible for the reader not familiar with group theory. Note that therefore some of the extra complications or subtleties are swept under the rug at the discretion of the author, especially if they are not believed to be relevant for understanding our model. Everything written below can be found in the following two references \cite{Dummit}, \cite{Herstein}.

\subsection*{Groups} A group is defined as the pair $(G, *_G)$, where $G$ is a set of elements and $*_G$ is a binary operation that satisfies the following three properties:

\begin{enumerate}
\item There exists an identity element in $G$ under $*_G$. That is, there is an element id$_G$ such that id$_G *_G x = x$ for all $x \in G$. 
\item The elements of $G$ are self contained under $*_G$. That is, for all $x, y \in G$, $(x *_G y) \in G$. 
\item There exists an inverse element for all elements of $G$. That is, for all $x \in G$, there exists $x^{-1}$ such that $x *_G x^{-1} = $ id$_G$. 
\end{enumerate}

An example of a group is the integers under addition. It is easily verified that all three conditions are met, where the identity element is $0$ and the inverse of $n$ is $-n$. 

For simplicity, groups will now be referred to as just $G$, where $*_G$ is implicitly assumed. 

\subsection*{Abelian} A group is said to be abelian if the elements of $G$ commute. That is, if for all $x, y \in G$, $x *_G y = y *_G x$. 

For the integers under addition, this is clearly the case. But most groups are not abelian (e.g. the set of matrices with unit eigenvalues under matrix multiplication are not abelian). 

\subsection*{Homomorphisms} A homomorphism is a map, $\varphi$, between two groups, $G$ and $H$, such that $\varphi(x *_G y) = \varphi(x) *_H \varphi(y)$ for all $x,y \in G$. Note that the binary operations are different on each side of the equation. 

In simpler terms, a homomorphism is a map that collapses one group onto another, while preserving some structure. For instance, the parity map (i.e. the map that returns $0$ if the argument is even and $1$ if the argument is odd), $\varphi$, from the integers to the integers modulus $2$ (i.e. $(\{0, 1\}, + \text{mod}(2))$, where $1 + 1= 0$ mod$(2)$) is a homomorphism. $\varphi$ clearly collapses the integers (it reduces them to a set with only two elements), but it preserves some structure (namely, parity). 

\subsection*{Free groups} A free group, $F(G)$, is an infinite group (in the sense that the set that comprises the elements of the free group is infinite) that is comprised of every possible combination of the elements of the set $G$, using the binary operation $*$. For instance, if $G = \{a,b, a^{-1}, b^{-1}\}$, then the set comprising $F(G)$ is given by $\{a, b, a * a, a * b, b * a, a^{-1} * b, a * b^{-1}, a^{-1} * b^{-1}, a * a * a, ...\}$. 

\subsection*{Acknowledgements}
\small We thank Sylvain Cappell for introducing us to free groups and for his clear explanation of the UFPG, Sanchit Chaturvedi and Roy Rinberg for their insightful discussions, and Nick Verga for inviting us to present our work early on. Finally, we thank Michael Berry and Dima Rinberg for discussing their work with us.

\end{document}